\documentstyle[preprint,aps]{revtex}   
\begin{document}
\draft

\title{
A fractal model for the first stages of thin film growth}

\author{Pablo Jensen (a)$^*$, Albert-L\'aszl\'o Barab\'asi (b), Hern\'an
Larralde (c), Shlomo Havlin (d) and H.~Eugene~Stanley (e)}

\address{ (a) D\'epartement de Physique des Mat\'eriaux, Universit\'e
Claude Bernard Lyon-1, Villeurbanne Cedex, France; (b) University of
Notre Dame, Department of Physics, Notre Dame, IN 46556, USA (c)
Physics Department, Cavendish Laboratory, Cambridge University,
Madingley Road, Cambridge CB3 0HE, England (d) Physics Department, Bar
Ilan University, Ramat Gan, Israel; (e) Center for Polymer Studies and
Department of Physics, Boston University, Boston, MA 02215, USA}

\maketitle

\begin{abstract}

In this paper we briefly review a model that describes the
diffusion-controlled aggregation exhibited by particles as they are
deposited on a surface.  This model allows to understand many
experiments of thin film deposition. In the first part, we describe
the model, which incorporates deposition, particle and cluster
diffusion, and aggregation. In a second part, we study the dynamical
evolution of the model. Finally, we analyze the effects of small
cluster mobility, and we show that the introduction of cluster
diffusion dramatically affects the dynamics of film growth. Some of
these effects can be tested experimentally.
\end{abstract}

\pacs{}
\narrowtext

\section{Introduction}
Understanding the processes underlying the growth of thin films has
led to widespread interest \cite{review}, both from theoretical
\cite{review,stoyanov,venables,tang,pubdda,bartelt,villain,bales} and
experimental points of view \cite{mo,noneq,roder,bardotti}.  Recently,
improvements in experimental techniques---such as scanning tunneling
microscopy---permit the investigation of atomic details of the
embryonic ``sub-monolayer'' stages of film growth. Recent experimental
works \cite{noneq,roder,bardotti} have shown that {\it fractal}
structures can be formed during the first stages of the growth. Then,
it is interesting to try to analyze the growth in the framework of
fractal models. One might consider the use of the percolation model
\cite{stauffer,bunde} to describe certain experiments of surface
deposition \cite{lecbd}.  However, percolation assumes that particles
do not diffuse after being deposited, when in fact not only diffusion
but also aggregation of the diffusing particles takes place.  There
exist models of diffusing particles that aggregate, but such
``cluster-cluster aggregation'' (CCA) models \cite{meakin} do not
allow the continual injection of new particles via deposition. Here we
develop a model that incorporates the three physical ingredients of
thin film growth: deposition, diffusion and aggregation (DDA).
Similar models that neglect the shape of the islands or the
possibility of cluster diffusion were studied independently
\cite{tang,bartelt}. We have shown in detail elsewhere \cite{pubdda}
how the DDA model generates a wide variety of fractal structures
characteristic of different models such as percolation, diffusion
limited aggregation (DLA) \cite{dla,vicsek} or CCA. Here we focus on
the dynamics of film growth and the importance of (small) island
diffusion.

\section{Model description and justification}

The DDA model is defined as follows (Fig.\ \ref{dda}):

(1) {\it Deposition}.  Particles are deposited at randomly-chosen
positions of the surface at a flux $F$ per lattice site per unit time.

(2) {\it Diffusion}.  All particles and clusters (sets of connected
particles) are chosen at random and attempted to move North, East, South
or West by one lattice constant per unit time. The probability that they
actually move is proportional to their mobility, which we assume to be
given by $D_s = D_1 s^{-\gamma}$.  Here $s$ is the number of particles
in the cluster, $D_1$ is the diffusion coefficient for a monomer
($s=1$), and the parameter $\gamma$ characterizes the dependence of
$D_s$ on cluster size.

(3) {\it Aggregation.} If two particles come to occupy neighboring
sites, they (and therefore the clusters to which they belong) stick
irreversibly.

We call {\it particles} the isolated atoms (or monomers) that are
deposited on the surface, {\it clusters} any set of connected
particles (including the monomers) and {\it islands} the clusters
containing more than one particle.  Physically, two competing
mechanisms are introduced in the model, each one with its own time
scale: deposition and diffusion.  It is useful to introduce the
normalized flux defined as the number of particles deposited per unit
site per {\it diffusion time\/} $\tau$, where $\tau$ is the mean time
needed by a monomer to jump by a lattice site. The monomer diffusion
coefficient is then given by $D_1=1/(4 \tau)$, and the normalized flux
by $\phi=F\tau$. Then, from experimental values of $F$ and $D_1$ it is
possible to calculate $\phi$ and the morphologies predicted by our
model.  The program actually calculates a probability for dropping a
particle: $p_{drop}=\phi L^2/(\phi L^2+N_{cl})$ where $L$ is the
system size and $N_{cl}$ is the total number of clusters present in
the system. A random number $p$ is chosen and compared to
$p_{drop}$. If $p < p_{drop}$, a particle is added at a random
position on the lattice. If $p > p_{drop}$, a cluster or a particle is
chosen at random and attempted to move. In both cases, the time is
increased by $\tau/(\phi L^2+N_{cl})$.

It should be stressed that this is only a ``zeroth-order'' model
which has the ambition to give a feeling on the relative influence of
deposition and diffusion on the growth properties of films. Details
specific to certain experimental systems, such as the existence of the
Schwoebel barrier, the precise dependence of cluster diffusion on
size, etc. are not carefully taken into account since we want to keep
the DDA model as general as possible (see Ref. \cite{pubdda}).

\section{Dynamical Evolution}

We present here snapshots of the system at different times to show the
formation of the islands. These images were obtained for
$\phi=1.2 \ 10^{-8}$ and $\gamma=\infty$ (i.e., only monomers are allowed
to move). A detailed presentation of the dynamical evolution of the
model has been given elsewhere \cite{pubdda}. Actually, it is more
interesting to use the surface coverage $\theta$ instead of the time
as the evolution parameter. $\theta$ is defined as the ratio of the
number of occupied sites to the total number of sites on the surface,
and for the times studied here we have $\theta \sim Ft$.

At very short times ($\theta\leq 0.001$), mainly monomers (isolated
particles) are found on the substrate, since they did not yet met
another one to form a cluster. Later, small clusters are homogeneously
grown on the surface (Fig.\ \ref{morph}a), and the island density
(i.e. the number of islands per lattice site) starts to grow. These
small clusters can be considered as the ``nucleation centers'' for the
growth.  As time increases, large clusters grow on these nucleation
centers, by addition of single particles (Fig.\ \ref{morph}b).  These
clusters are very similar to those obtained experimentally
(Figs. \ \ref{fractal}a-b). We have found that their fractal dimension
is 1.65, very close to the fractal dimension of the DLA clusters.
This (DLA-like) growth mechanism goes on until the linear dimension of
the clusters becomes comparable to the separation between them (Fig.\
\ref{morph}c). Then, many particles start to fall {\it inside\/} the
clusters and their fractal dimension rapidly increases. Eventually a
cluster of a size comparable to the system size is built and the
system spans (Fig.\ \ref{morph}d).  At that time, the effective
fractal dimension of the spanning cluster reaches a value close to
1.9.

It is interesting to note that the growth of the film for
$\gamma=\infty$ is in some loose sense "self-regulated" since islands
tend to avoid each other, by growing preferentially in the directions
where no island is present. The reason is that monomers are more
abundant in those regions since no island captures them. Then, the
probability for a given point of an island to grow is higher if it is
far from other islands. This effect can be quantified in the following
way. We first grow a system with a given coverage and we freeze the
growth. Then, we deposit a monomer and we move it exactly as during
the growth, but each time the monomer reaches an island, we record the
point where it touched the island and {\it remove} it. By counting the
number of monomers that hit each point, we can measure the growth
probability within each island. Fig.\ \ref{proba}a shows a global view
of these probabilities for a large system at a coverage of \ .04. The
white points correspond to those point which grow slower and the
bright blue points show the regions of high growth probability. Fig.\
\ref{proba}b shows in detail the growth probabilities for one of the
islands. The fact that each island has its own ``capture zone'' and
grows inside it has some interesting consequences on the size
distribution of the islands. These will be discussed below.

\section{Island diffusion}

In the last paragraph we have studied the behavior of the model in the
case when only monomers are allowed to move. In this section, we want
to address some consequences of (small) cluster diffusion. It has long
been recognized that cluster diffusion can influence the growth of the
films \cite{stoyanov,cludiff}, even if at that time experimental
proofs were lacking. Recently, experiments \cite{cludifexp} and
molecular dynamics studies have shown that small clusters can move on
the surfaces without breaking \cite{depristo}. Generally, however, the
experimental results are analyzed in the framework of the different
models containing only monomer diffusion. These models can include
{\it reversible} aggregation \cite{bales,smilauer}, meaning that two
monomers that aggregate can detach after a certain time. This can lead
to wrong interpretations if small cluster mobility is indeed present in
the experiments \cite{metiu}. Then, it is important to understand the
consequences of this mobility to be able to identify them in the
experiments.

We have shown previously \cite{pubdda} that the introduction of
cluster mobility considerably changes the growth dynamics of the
film. Specifically, we showed that:

({\it i\/}) The mean cluster size increases {\it exponentially} as a
function of the coverage. A power-law dependence is generally found
when only monomer diffusion is allowed. This effect is intriguing and
may be due to large cluster diffusion \cite{pubdda}.

({\it ii\/}) The maximum island density depends on the incident flux
according to $N_{max} \sim F^\zeta$ with $\zeta=0.42$, as opposed to
$\zeta=0.33$ found with monomer diffusion. It is also known that when
only {\it dimer} diffusion is introduced, $\zeta=0.4$
\cite{villain}. The problem is that generally the accuracy of the
experimental determinations of $N_{max}$ is not high enough to
distinguish between the different exponents
\cite{bardotti}.

({\it iii\/}) More interesting experimentally: the evolution of the
island density as a function of coverage for low normalized fluxes
(typically less than $10^{-5}$). It is known \cite{tang} that the
maximum of the island density is reached only at roughly a coverage of
0.2 when only monomers can move.  On the contrary, if clusters can
move, this maximum is reached for {\it smaller} values of the
coverage.

We add here a test that is easy to perform experimentally, and which
has actually already been used to distinguish different growth
mechanisms \cite{bardotti,metiu,stroscio}: the rescaling of island
size distributions \cite{bales,rescale}. The idea is to rescale the
island size distributions into a universal distribution that depends
on the ingredients of the model (i.e. the detailed mechanisms of
Deposition, Diffusion and Aggregation) but not on the values of the
flux or the coverage. Very recently, Mulheran and Blackman
\cite{mulheran} have given some interesting insights of why such a
universal function should exist, at least for the case of
heterogeneous growth. Their argument can be summarized as follows. At
the beginning of the growth, nucleation centers form (Fig.\
\ref{morph}a). Then, each center grows by catching the monomers
falling inside its ``capture zone'', roughly identified with its
Vorono\"{\i} polyhedron. Therefore its size is, at any time, proportional
to the surface of its Vorono\"{\i} polyhedron, which does not change with
time (if one neglects nucleation of new islands in the case of
homogeneous nucleation). The result is that at any coverage the size
distribution of the islands reproduces that of the Vorono\"{\i} cells,
which explains the rescaling for different coverages. We show in
Figure\ \ref{distrib} that the scaling of the size distributions also
occurs when clusters are allowed to move, but the universal function
is different from that found when only monomers move. Three different
diffusion hypothesis have been made: only monomers move, monomers and
dimers diffuse, all clusters up to size 100 do move ($\gamma=1$ has
been taken in the two last cases). We see that the universal function
becomes narrower and has a higher maximum when larger and larger
clusters are allowed to diffuse.

\section{Discussion and Conclusion}
In summary, we have proposed a model that describes the
diffusion-controlled aggregation exhibited by particles as they are
deposited on a surface.  The model, which incorporates deposition,
particle and cluster diffusion, and aggregation closely reproduces
some experimental images (compare Figs\ \ref{fractal}a,b and Figs\
\ref{morph}). We find that the model permits one to distinguish the
effects of deposition, diffusion and aggregation, and that tuning the
relative strength of, e.g., deposition and diffusion, generates a rich
range of morphologies---including diffusion limited aggregation,
cluster-cluster aggregation, and percolation.  The length and time
scales characterizing these morphologies depend on such
experimentally-controllable parameters as deposition flux and
diffusion constant, raising the possibility that the model may prove
useful in future studies seeking the controlled design of
nanostructure morphologies.  We can argue that the DDA model is suited
for ({\it i}) MBE by taking large $\gamma$ values.  ({\it ii}) Other
deposition experiments where edge diffusion is absent, for example
aggregate deposition \cite{bardotti}.  ({\it iii}) All the dynamic
properties related to {\it small} cluster diffusion on surfaces such
as those presented in Sec. VI.

PJ acknowledges CNRS for financial support.  HL thanks CONACYT,
Mexico, for support.  A-LB and HES thank the Hungary-USA exchange
program of the Hungarian Academy of Sciences.  SH acknowledge the
USA-Israel Binational Science Foundation for financial support. The
Center for Polymer Studies is supported by NSF.

\begin{figure}
\caption{
Experimental images of the first stages of growth for films prepared
by (a) Atomic deposition of Ag on Pd \protect \cite{roder} and
(b) Deposition of Sb {\it compact} aggregates containing 2300 atoms on
graphite \protect \cite{bardotti}.}
\label{fractal}
\end{figure}

\begin{figure}
\caption{Schematic representation of the basic processes considered in
this model: (a) deposition, (b) and (d) particle diffusion, (e) island
diffusion and (c) aggregation. (b) corresponds to {\it nucleation}
(i.e.  a new island (c) is created) while (d) corresponds to {\it
growth} of an already existing island (see the text for details).}
\label{dda}
\end{figure}

\begin{figure}
\caption{
Morphologies obtained for a normalized flux $\phi=1.2 \ 10^{-8}$, a
system size $L=500$ and $\gamma=\infty$. The images show a portion 290 x
210 of the lattice. Four different coverages (corresponding to four
different times of deposition) are shown: (a) coverage 0.02 (b)
coverage 0.1 (c) coverage 0.25 (d) coverage 0.4. Colors indicate the
order in which particles were deposited on the surface: the first
particles deposited are colored white, then yellow, orange, red, blue
and finally green.}
\label{morph}
\end{figure}

\begin{figure}
\caption{
Growth probabilities for different points of the islands at a coverage
0.04 ($\gamma=\infty$). (a) global view of the surface (b) detailed view
of an island. Colors give the probability that a given site of the
island has to catch a monomer. The probability increases in the order:
white, yellow, orange, red, blue, green and finally big bright blue
dots.}
\label{proba}
\end{figure}

\begin{figure}
\caption{
Rescaled island size distributions. The island distributions are
transformed to p(s) which represents the probability that a randomly
chosen cluster belongs to an island containing s clusters
\protect\cite{bales}. The island size s is scaled by the mean island
size $s_m$. Sets (1) to (3) correspond to simulations where only
monomers are allowed to move ($\gamma=\infty$). For sets (4) to (6),
$\gamma=1$ but only dimers can move, while for sets (7) and (8)
clusters containing up to 100 monomers can move. The detailed
parameters used for each set are as follows: (1) $\phi=10^{-8}$,
coverage 0.3 ;(2) $\phi=10^{-8}$, coverage 0.15 ; (3) $\phi=10^{-9}$,
coverage 0.1 ; (4) $\phi=10^{-9}$, coverage 0.05 ; (5) $\phi=10^{-9}$,
coverage 0.1 ; (6) $\phi=10^{-8}$, coverage 0.3 ; (7) $\phi=10^{-8}$,
coverage 0.15 ; (8) $\phi=10^{-8}$, coverage 0.05 ; (9)
$\phi=10^{-8}$, coverage 0.1. The lines represent averages of the
distributions for each case.}
\label{distrib}
\end{figure}

\end{document}